\documentclass[%
aip,
amsmath,amssymb,
reprint,%
]{revtex4-1}

\usepackage{graphicx}% Include figure files
\usepackage{dcolumn}% Align table columns on decimal point
\usepackage{bm}% bold math
\usepackage[utf8]{inputenc}
\usepackage[T1]{fontenc}
\usepackage{mathptmx}
\usepackage{booktabs}
\usepackage{multirow}
\usepackage{gensymb}
\usepackage{mathrsfs}

\begin{document}
	
\preprint{AIP/123-QED}
	
\title{Influence of clamp-widening on the quality factor of nanomechanical silicon nitride resonators}
	
\author{Pedram Sadeghi}
\affiliation{Institute of Sensor and Actuator Systems, TU Wien, Gusshausstrasse 27-29, 1040 Vienna, Austria}
	
\author{Manuel Tanzer}
\affiliation{Institute of Sensor and Actuator Systems, TU Wien, Gusshausstrasse 27-29, 1040 Vienna, Austria}
	
\author{Simon L. Christensen}
\affiliation{DTU Fotonik, Department of Photonics Engineering, Technical University of Denmark, {\O}rsteds Plads 343, DK-2800 Kongens Lyngby, Denmark}
	
\author{Silvan Schmid}
\email{silvan.schmid@tuwien.ac.at}
\affiliation{Institute of Sensor and Actuator Systems, TU Wien, Gusshausstrasse 27-29, 1040 Vienna, Austria}
	
\date{\today}% It is always \today, today,
%  but any date may be explicitly specified
	
\begin{abstract}
	Nanomechanical resonators based on strained silicon nitride (Si$_3$N$_4$) have received a large amount of attention in fields such as sensing and quantum optomechanics due to their exceptionally high quality factors ($Q$s). Room-temperature $Q$s approaching 1 billion are now in reach by means of phononic crystals (soft-clamping) and strain engineering. Despite great progress in enhancing $Q$s, difficulties in fabrication of soft-clamped samples limits their implementation into actual devices. An alternative means of achieving ultra-high $Q$s was shown using trampoline resonators with engineered clamps, which serves to localize the stress to the center of the resonator, while minimizing stress at the clamping. The effectiveness of this approach has since come into question from recent studies employing string resonators with clamp-tapering. Here, we investigate this idea using nanomechanical string resonators with engineered clampings similar to those presented for trampolines. Importantly, the effect of orienting the strings diagonally or perpendicularly with respect to the silicon frame is investigated. It is found that increasing the clamp width for diagonal strings slightly increases the $Q$s of the fundamental out-of-plane mode at small radii, while perpendicular strings only deteriorate with increasing clamp width. Measured $Q$s agree well with finite element method simulations even for higher-order resonances. The small increase cannot account for previously reported $Q$s of trampoline resonators. Instead, we propose the effect to be intrinsic and related to surface and radiation losses.
\end{abstract}
	
\maketitle
	
\section{\label{sec:introduction}Introduction\protect}
The search for nanomechanical resonators with increasingly high quality factors ($Q$s) has grown into a field of its own in the last decade \cite{Ekinci2005}. Increasing $Q$ e.g. directly reduces the intrinsic force noise \cite{Cleland2002}, which enables force sensors with attonewton sensitivity \cite{Stowe1997}, allowing the detection of single electron spins \cite{Rugar2004}. For resonant force sensing based on the vibrational amplitude, larger $Q$s directly increase the force responsivity defined as $\mathscr{R} = \frac{Q}{k}$, with $k$ being the spring constant of the resonator \cite{Schmid2016}.
	
In the field of quantum optomechanics, the $Q \times f$ product, where $f$ is the resonance frequency of the mechanical resonator, is often used to quantify the decoupling of a resonator from a thermal bath \cite{Aspelmeyer2014}. The requirement for performing quantum experiments at room-temperature is then calculated to be $Q \times f > 6 \times 10^{12}$ Hz. A high-$Q$ nanomechanical device would be optimal due to the additional large frequencies associated with scaling down resonator size.
	
A general trend is that the $Q$ decreases with decreasing dimensions of the resonator, attributed to losses at the resonator surface due to the increased surface-to-volume ratio \cite{Carr1999}. However, it was discovered that highly stressed silicon nitride (Si$_3$N$_4$) resonators achieve $Q$s on the order of a million at room-temperature \cite{Verbridge2006,Verbridge2008}. Through stress engineering, it was found that the large $Q$s are a direct result of the added tension \cite{Verbridge2007}. The observed $Q$-enhancement has been attributed to dissipation dilution, where the tensile stress dilutes the intrinsic losses of the material \cite{Gonzalez1994,Huang1998,Unterreithmeier2010,Schmid2011}. Dissipation dilution has also been observed in other materials under tensile stress \cite{Schmid2008,Cole2014} and is a universal effect of strained resonators \cite{Cagnoli2000,Fedorov2019}.
	
Acoustic radiation losses remain a major loss mechanism, in particular in high-stress nanomechanical resonators. Introduction of phononic crystal structures has circumvented this problem by suppressing the tunnelling of phonons into the substrate \cite{Yu2014,Tsaturyan2014}. Engineering the phononic crystal directly into the resonator additionally reduces bending at the clamping, creating soft-clamping, successfully demonstrated in both membrane \cite{Tsaturyan2017} and string \cite{Ghadimi2017} resonators. Combining soft-clamping with strain engineering, where the stress in the resonator is increased several-fold from the uniform case, $Q \times f > 1 \times 10^{15}$ Hz has been achieved at room-temperature \cite{Ghadimi2018}. Currently, surface losses remain the fundamental limit in all the approaches presented above \cite{Villanueva2014}.
	
While the regularly-clamped and soft-clamped resonators can be understood in the framework of dissipation dilution, Norte et al. observed anomalously large $Q$s in nanomechanical trampoline resonators made of Si$_3$N$_4$ \cite{Norte2016}. Values approaching the order of $10^8$ were observed for the fundamental out-of-plane modes of trampolines of diagonal length $L \approx 1$ mm and thickness $h = 20$ nm, more than one order of magnitude larger than expected for a string resonator of similar dimensions. The large $Q$s were attributed to curved widening of the trampoline clamping to widths larger than the width of the tethers. This clamp-widening served to minimize the stress in the clamping, while enhancing the stress in the tethers to values near the yield strength of Si$_3$N$_4$. These results would present a much simpler way of achieving ultra-high $Q$s compared to the aforementioned soft-clamping and strain engineering approaches, which are highly fragile structures and challenging to fabricate.
	
Additional studies have been made employing the same trampoline geometry for ultralow-noise sensors \cite{Reinhardt2016} and magnetic resonance force microscopy \cite{Fischer2019}. While large $Q$s are observed in these cases as well, they can be explained through the framework of dissipation dilution. A recent study by Bereyhi et al. investigated the effect of varying the width at the clamping in string resonators \cite{Bereyhi2019}. Contrary to the investigations on trampoline resonators, the $Q$ of the fundamental mode deteriorated with increasing clamp width. Instead, the $Q$ was increased by more than a factor of two for clamp-tapered strings. Theoretical models of dissipation dilution based on the mode-shape of the resonance agreed well with the measured data. The idea that $Q$s of trampoline resonators are enhanced as a result of clamp-widening was therefore put into question. However, while in the initial work on silicon nitride trampoline resonators \cite{Norte2016}, the tethers featuring the widening were oriented diagonally with respect to the supporting silicon frame, in the investigation performed by Bereyhi et al. they were oriented perpendicular to the frame. 
	
In this report, we investigate the effect of clamp-widening on string resonators that are oriented both diagonally and perpendicularly with respect to the silicon frame. Using low-stress Si$_3$N$_4$, a comparison is made between $Q$s in the two clamping configurations for increasing clamp radius and the results are compared to finite element method (FEM) simulations of dissipation dilution. It is found that the $Q$ is slightly increased in the diagonal configuration for small clamp radii, but drops for larger radii. In the perpendicular case, $Q$ steadily falls with increasing clamp radius, similar to what was reported by Bereyhi et al. Additionally, the dependence of $Q$ on the mode number is investigated and the results found to agree well with FEM simulations. These results suggest that only slight enhancement of the $Q$ is expected for clamp-widened strings oriented in a diagonal fashion. Alternative possible explanations are given with a focus on surface and radiation losses.
	
\section{\label{sec:methods}Methods\protect}
The fabrication process of the nanomechanical string resonators is similar to previous studies \cite{Schmid2010,Luhmann2017}. Samples are defined on commercial silicon wafers coated with 50 nm Si$_3$N$_4$ by low pressure chemical vapor deposition. String designs are defined in the Si$_3$N$_4$ on the topside by UV lithography followed by a dry etching step. Backside openings are similarly defined and the strings released by means of an anisotropic KOH (40 wt. $\%$) wet etching at 80 $\degree$C.
	
The mechanical properties of the resonators are measured optically using a commercial laser-Doppler vibrometer (MSA-500 from Polytec GmbH). A secondary diode laser (LPS-635-FC from Thorlabs GmbH) is amplitude modulated to actuate the vibrations, either by focusing the spot to the anti-node of a resonance (radiation pressure) or to the rim of the string (thermoelastic). All measurements are performed in high vacuum (pressure p $\sim 10^{-5}$ mbar) to minimize gas damping \cite{Schmid2011_2}.
	
$Q$s are extracted by feeding the analog signal of the vibrometer to a lock-in amplifier (HF2LI from Zurich Instruments). The lock-in amplifier allows ring-down measurements to be performed, which is done by actuating the sample at resonance, locking the mode with a phase-locked loop (PLL), and then turning off the drive.
	
\begin{table}[ht!]\centering
	\begin{tabular}{lcc}
		\toprule
		& \multicolumn{2}{c}{Clamping configuration}\\
		\midrule
		Assumed material parameters\hspace{30pt} & Diagonal\hspace{10pt} & Perpendicular \\
		\midrule
		Young's modulus ($E$) & \multicolumn{2}{c}{250 GPa}\\
		Mass density ($\rho$) & \multicolumn{2}{c}{3000 kg/m$^{3}$}\\
		Poisson's ratio ($\nu$) & \multicolumn{2}{c}{0.23}\\
		String central width & \multicolumn{2}{c}{5 $\mu$m}\\
		String thickness (h) & 50 nm & 56 nm\\
		Tensile prestress ($\sigma_p$) & 0.14 GPa & 0.47 GPa\\
		\bottomrule
	\end{tabular}
	\caption[Table 1]{Material parameters used in the FEM simulations for both configurations. Geometric dimensions and tensile stress are measured for each sample, while the rest are assumptions based on previous literature on the properties of Si$_3$N$_4$.}\label{tab:1}
\end{table}
	
For uniform string resonators, $Q$s can be calculated analytically from the energies of the system \cite{Unterreithmeier2010}. For small amplitudes, the $Q$ due to dissipation dilution of a string can be calculated from \cite{Gonzalez1994,Huang1998,Yu2012,Schmid2016}:
\begin{equation}\label{eq:1}
	Q = \left[\frac{\left(n \pi\right)^2}{12} \frac{E}{\sigma} \left(\frac{h}{L}\right)^2+ \frac{1}{\sqrt{3}} \sqrt{\frac{E}{\sigma}} \frac{h}{L}\right]^{-1} Q_{\text{int}},
\end{equation}
where $n$ is the mode number, $E$ is Young's modulus, $\sigma$ is the relaxed tensile stress in the resonator, $L$ is the length, and $h$ is the thickness. Values of fixed parameters in this study are shown in Table \ref{tab:1}. $Q_{\text{int}}$ is the intrinsic quality factor, defined as \cite{Villanueva2014}:
\begin{equation}\label{eq:2}
	Q_{\text{int}}^{-1} = Q_{\text{mat}}^{-1} + Q_{\text{surf}}^{-1}
\end{equation}
where $Q_{\text{mat}} \approx 28000 \pm 2000$ is the bulk material loss of Si$_3$N$_4$, and $Q_{\text{surf}} = \beta \cdot h$ the surface loss with $\beta = 6 \times 10^{10}\pm 4 \times 10^{10}$ $\text{m}^{-1}$. For the samples studied here with thicknesses $h \approx 50$ nm, $Q_{\text{surf}} \approx 3000$ and should therefore dominate the intrinsic losses.
	
Because of the non-trivial string geometries, measured $Q$s cannot readily be compared to calculations using Equation \ref{eq:1} for clamp-widened strings. Therefore, we employ finite element method simulations using COMSOL Multiphysics in order to simulate the mode shapes. A shell interface of the structural mechanics module is employed due to the large aspect-ratios of the strings. From Rayleigh's method, we can assume the maximum kinetic energy equals the total stored energy of the system, $W_{\text{kin}}^{\text{max}} = W_{\text{stored}}$, while the energy stored in the bending is set equal to the total strain energy, $W_{\text{strain}}$ \cite{Schmid2016}. The dissipation dilution factor is then calculated directly from the mode-shape and given as $\alpha_{\text{DD}} = \frac{W_{\text{kin}}^{\text{max}}}{W_{\text{strain}}}$. Setting $Q_{\text{int}}$ as a fit parameter, the quality factor can be calculated from $Q = \alpha_{\text{DD}} Q_{\text{int}}$.
	
\begin{figure*}[ht!]
	\centering
	\includegraphics[width=\textwidth]{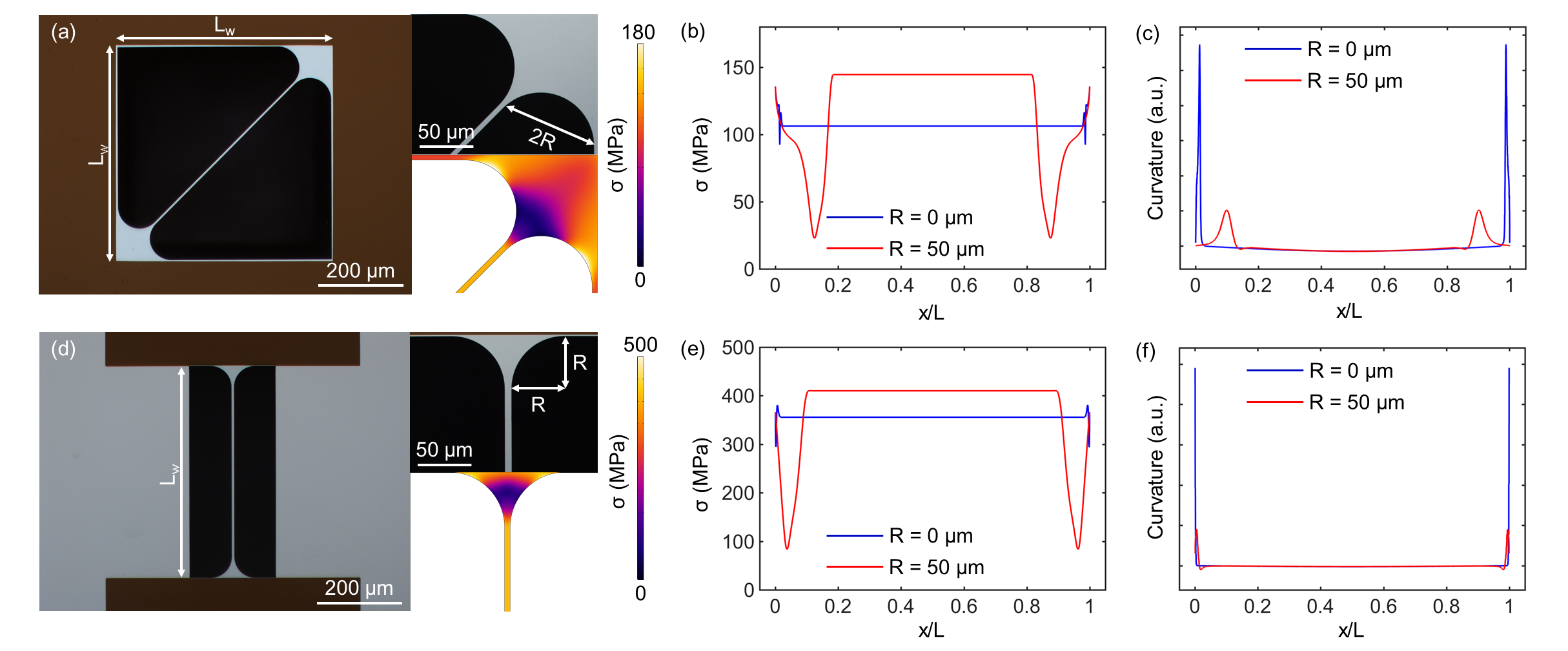}
	%  \vspace*{-28mm}
	\caption[Figure 1]{Clamp-widened string geometries investigated here. (a) Optical micrograph of a diagonally oriented string with a 500 $\mu$m wide opening window. A zoom-in on the clamping region is also shown including a stress profile. (b) Stress profile along the length of the diagonal string. (c) Curvature along the length of the diagonal string. (d) Optical micrograph of the perpendicular string geometry for a string with $L_w = 500$ $\mu$m. (e) Stress profile along the length of the perpendicular string. (f) Curvature along the length of the perpendicular string.}\label{fig:1}
\end{figure*}
	
\section{\label{sec:results}Results and Discussion\protect}
Figure \ref{fig:1} shows optical micrographs and simulated stress distributions for the two clamping configurations investigated. The string can be anchored in either a diagonal or perpendicular configuration, as shown in Figures~\ref{fig:1}a\&d, respectively. The diagonal design is inspired from previous investigations on trampoline resonators \cite{Norte2016,Reinhardt2016}, while the perpendicular clamping configuration is similar to the work of Bereyhi et al. \cite{Bereyhi2019}. However, while Bereyhi et al. had a constant clamp-widening separated by a transition region from the central string, our clamping design only consists of a round transition region of increasing width.
	
Figures~\ref{fig:1}a\&d also show a zoom-in on the clamping region, highlighting how the clamp-widening is defined through the radius of the clamping region, $R$, in either case. Low-stress Si$_3$N$_4$ is used for the investigation, since lower stress increases the ratio of phase velocities (acoustic mismatch) between resonator and substrate \cite{Wilson-Rae2011,Villanueva2014}. As a result of different batches of Si$_3$N$_4$ being used for the two configurations, the prestress in the diagonal configuration is $\sigma_p \approx 0.14$ GPa, while the perpendicular configuration has $\sigma_p \approx 0.47$ GPa.
	
\begin{figure*}[ht!]
	\centering
	\includegraphics[width=\textwidth]{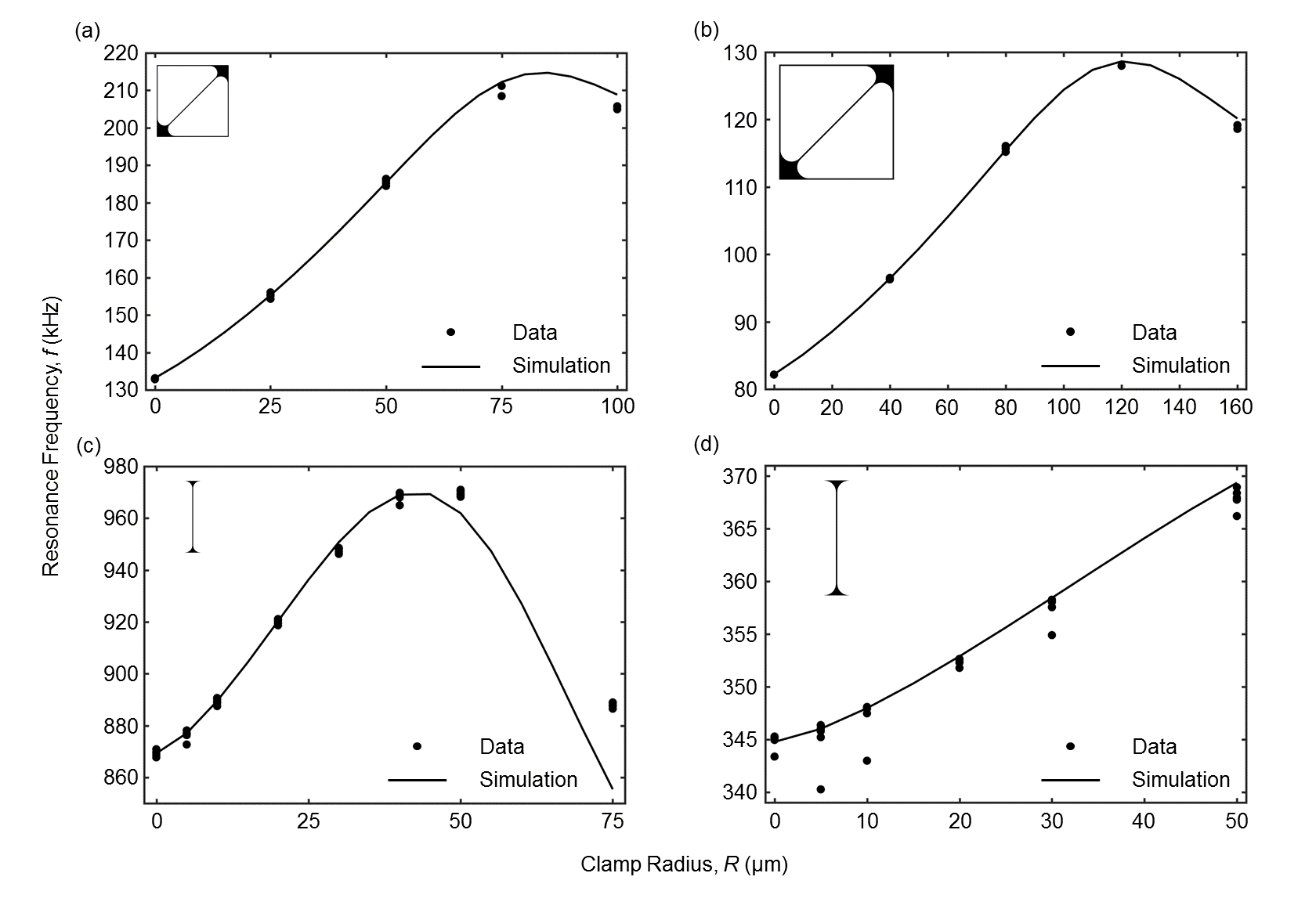}
	%  \vspace*{-28mm}
	\caption[Figure 2]{Resonance frequencies of the fundamental out-of-plane mode of clamp-widened string resonators. Measured resonance frequencies of diagonal strings with a window size (a) $L_w = 500$ $\mu$m and (b) $L_w = 800$ $\mu$m as a function of increasing clamp radius $R$. Each data point corresponds to an individual sample. The solid lines represents FEM simulations. Resonance frequencies of strings in the perpendicular configuration are shown with (c) $L_w = 200$ $\mu$m and (d) $L_w = 500$ $\mu$m. Schematics on each plot are made for clarity.}\label{fig:2}
\end{figure*}
	
The zoom-ins presented in Figures~\ref{fig:1}a\&d also show the simulated stress profiles in either configuration. In both cases, the tensile stress is localized in the central, uniform part of the string, while the clamping region just outside the uniform region displays a much lower stress. The stress profile along the length of the diagonally oriented string is shown in Figure~\ref{fig:1}b, showing a minimum stress of only 16$\%$ the value in the string center. A similar profile for the perpendicular case is shown in Figure~\ref{fig:1}e. Both cases are compared to similarly clamped strings without clamp-widening (uniform strings), where the stress profile changes little with position along the string length. As expected, the stress is enhanced in the string center for the widened strings, similar to previous research on strain engineered strings \cite{Ghadimi2018}. Reduced stress at the clamping of widened strings additionally reduces bending at the clamping, which can be observed in Figures~\ref{fig:1}c\&f for the diagonal and perpendicular configurations, respectively. The maximum curvature with clamp-widening is $\sim$18$\%$ of the value without widening in both cases.
	
As has been shown, dissipation dilution can be optimized by either increasing the tensile stress or reduce bending at the clamping (soft-clamping). From the FEM analysis in Figure~\ref{fig:1}, strings with clamp-widening show both an increased stress in the string and reduced curvature at the clamping. These features would predict an increase of $Q$ in clamp-widened strings. However, while the string experiences a quasi soft-clamping, there is significantly more material that is bending at the clamping. The additional damping caused by the material used for clamp-widening will counteract the positive effects of increased stress and reduced curvature. Hence, the effect of clamp-widening has to be studied both by experiments and simulations.
	
Tensile prestress can be estimated by first finding the relaxed stress in the resonator using the analytical model for the eigenfrequency of a string:
\begin{equation}\label{eq:3}
	f = \frac{n}{2 L}\sqrt{\frac{\sigma}{\rho}}.
\end{equation}
The biaxial prestress is then calculated using $\sigma_p = \frac{\sigma}{1 - \nu}$. This value can be used for the simulation of clamp-widened strings. Confirmation of this approach is achieved by comparing measured and simulated frequencies as shown in Figure~\ref{fig:2} for the fundamental out-of-plane mode. For each configuration, two different lengths are displayed and the dependence on increasing radius is shown and compared to the case without any clamp-widening ($R = 0$ $\mu$m). Figures~\ref{fig:2}a\&b show the measured resonance frequencies of diagonally oriented strings with window lengths, $L_w$, of 500 $\mu$m and 800 $\mu$m, respectively, for increasing clamp radius. Measurements are compared with FEM simulations using the calculated prestress. Good agreement is achieved for both lengths, showing that the system is described well through the FEM simulations. Similar agreement is achieved for perpendicularly oriented strings as shown in Figures~\ref{fig:2}c\&d for a $L_w$ of 200  $\mu$m and 500 $\mu$m, respectively.
	
\begin{figure*}[ht!]
	\centering
	\includegraphics[width=\textwidth]{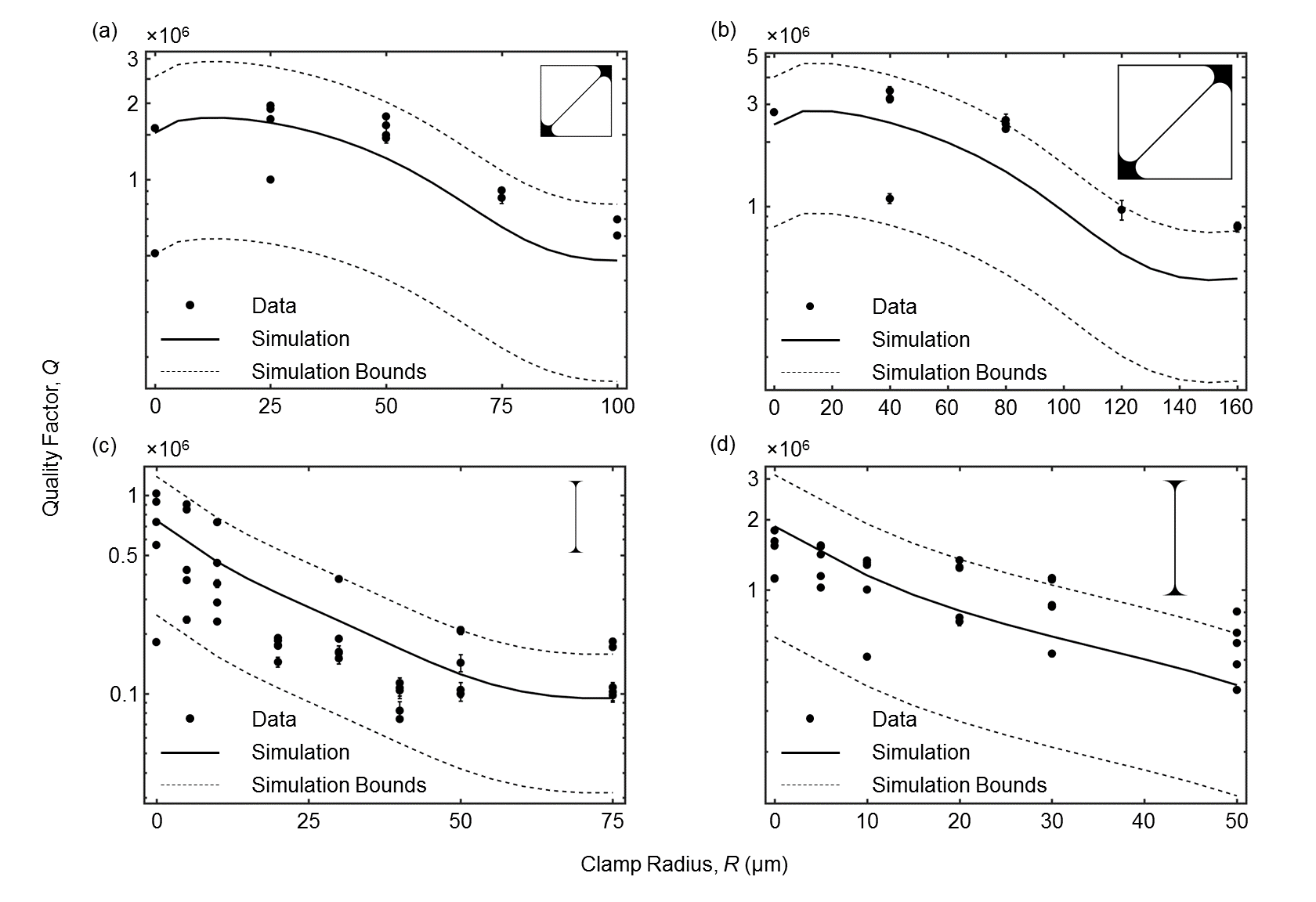}
	%  \vspace*{-28mm}
	\caption[Figure 3]{Quality factors of the fundamental out-of-plane mode of clamp-widened string resonators. (a) $Q$s of diagonal strings with a window size $L_w = 500$ $\mu$m for increasing clamp radius. Each data point is an average of 5 measurements of the same $Q$, while different data points correspond to different individual samples. The solid lines represent FEM simulations for the mean and the dashed lines the uncertainty of $Q_{\text{int}}= h \cdot 6\times 10^{10}\pm 4\times 10^{10}$ \cite{Villanueva2014}. (b) Same as (a) for $L_w = 800$ $\mu$m. (c) $Q$s of perpendicular strings with $L_w = 200$ $\mu$m. (d) Same as (c) for $L_w = 500$ $\mu$m. Schematics on each plots are added for clarity.}\label{fig:3}
\end{figure*}
	
Experimental measurements of the $Q$ for the fundamental out-of-plane mode in the two configurations are shown in Figure~\ref{fig:3}. FEM simulations of the $R$-dependence is given as well. Figure~\ref{fig:3}a shows measurements for diagonally oriented strings with a window size $L_w = 500$ $\mu$m. A slight increase of up to $\sim$23$\%$ in $Q$ is observed for $R = 25$ $\mu$m compared to $R = 0$ $\mu$m. For larger radii, the $Q$s start to roll off until a minimum is reached for $R = 100$ $\mu$m. The trend shows good agreement with the simulations. A similar observation is made for diagonally oriented strings with $L_w = 800$ $\mu$m (Figure~\ref{fig:3}b). The $Q$ peaks for $R = 40$ $\mu$m with a maximum $Q$ enhancement of $\sim$26$\%$ compared to $R = 0$ $\mu$m. For both lengths, the increase in $Q$ can be qualitatively predicted from the simulations. The scattering of the experimental $Q$ data is probably due to non-negligible radiation losses, sample imperfections, and variations of magnitude of the intrinsic damping.
	
Measured $Q$s in the perpendicular configuration are shown in Figure~\ref{fig:3}c\&d for strings with $L_w = 200$ $\mu$m and $L_w = 500$ $\mu$m, respectively. Unlike the diagonal case, $Q$s only get worse with increasing $R$. Simulations agree well with the observed trend and the spread in $Q$s for a given $R$ is well-accounted for by including the uncertainty in $Q_{\text{int}}$.
	
The results of the perpendicular configuration agree with the findings of Bereyhi et al., namely that the $Q$ steadily deteriorates with increasing clamp width \cite{Bereyhi2019}. Contrary to this observation, the diagonal configuration does display an increase in $Q$ for small radii. The underlying mechanism for this observation is not entirely understood, but is probably a result of different strain distributions at the clamping site for the two different designs. On the one hand, the level of $Q$ enhancement observed here cannot explain the observations made by Norte et al. on trampoline resonators \cite{Norte2016}. On the other hand, it was shown by Norte et al. that increasing the clamp radius caused an increase in $Q$, which contradicts the decrease observed here for larger radii. It should, however, be noted that the maximum radius shown by Norte et al. is much smaller than the maximum here, meaning that a similar trend might be observed for the trampolines if larger radii are designed as well.
	
\begin{figure*}[ht!]
	\centering
	\includegraphics[width=\textwidth]{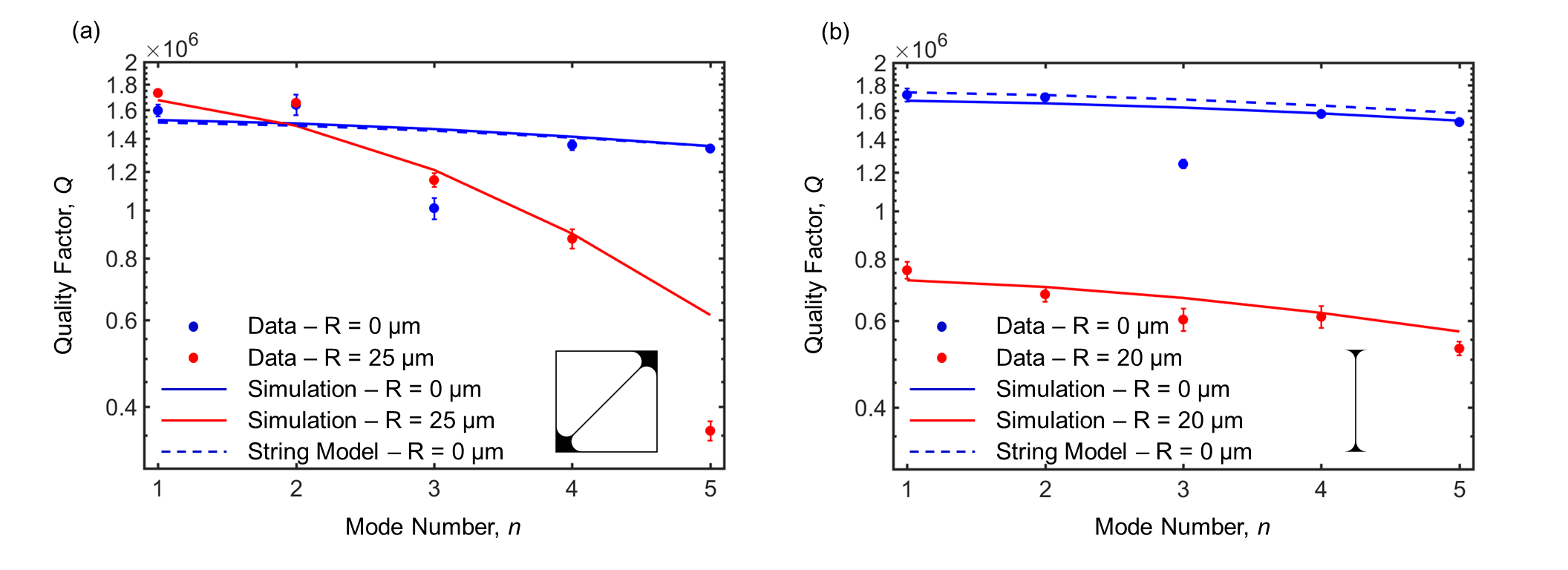}
	%  \vspace*{-28mm}
	\caption[Figure 4]{Quality factors of the out-of-plane vibrations of clamp-widened strings for higher order modes. (a) $Q$s for diagonal strings with $L_w = 500$ $\mu$m versus mode number without (blue points) and with (red points) a widened clamping with $R = 25$ $\mu$m. Each data point is an average of 5 measurements of the same resonance. Solid lines are FEM simulations and the dashed line analytical calculations using equation \ref{eq:1} for $R = 0$ $\mu$m. (b) Same as (a) for perpendicular strings with $L_w = 500$ $\mu$m and with the clamp-widened string having $R = 20$ $\mu$m. In both cases, $Q_{\text{int}} = 3000$ in the theoretical calculations. Schematics on each plots are made for clarity.}\label{fig:4}
\end{figure*}
	
To further investigate the clamp-widened strings, the $Q$ for higher order modes are studied, as shown in Figure~\ref{fig:4}. The measurements were made on strings with $L_w = 500$ $\mu$m in both configurations. A radius of $R = 25$ $\mu$m and $R = 20$ $\mu$m is chosen for the diagonal and perpendicular configuration, respectively. For larger radii, the clamping region starts to affect the mode-shape of the string significantly, making it hard to find higher-order modes.
	
Figure~\ref{fig:4}a shows the $Q$s of the first five out-of-plane modes of a diagonally oriented string with and without a widened clamping. FEM simulations of the $Q$s are shown as well for $Q_{\text{int}} = 3000$. In addition, $Q$s calculated using Equation~\ref{eq:1} are shown for the uniform strings. A $Q$-enhancement is only observed for the fundamental mode, while the $Q$ starts to deteriorate for higher order modes with a widened clamp compared to without widening. Mode-dependence for strings in the perpendicular configuration are shown in Figure~\ref{fig:4}b. A nominal $Q_{\text{int}} = 3000$ gives a good fit with experimental data points. In this configuration, $Q$s are lower with clamp-widening than without regardless of the mode number, as expected from data shown in Figure~\ref{fig:3}. Note that the $Q$ rolls off slower than for the diagonal strings.
	
The mode-dependence shows reasonable agreement with simulations for both configurations. For samples without clamp-widening, simulated $Q$s also agree very well with those calculated using the dissipation dilution model. As such, FEM is a valuable tool for predicting and designing $Q$s of strained resonators with complex geometries. Since a value of $Q_{\text{int}} = 3000$ provides a good fit with measurements, it can be concluded that none of the two samples are limited by radiation loss. In general, the intrinsic losses can be accounted for as a fit parameter. Despite the large spread in $Q$-values for each $R$, the mode-dependence matches simulations as long as $Q_{\text{int}}$ is adjusted accordingly.
	
From the investigation of the clamp-widened string samples presented above, it can be concluded that widening in the diagonal configuration results in slight $Q$ enhancements for the fundamental out-of-plane mode and cannot explain the large $Q$s observed for trampoline resonators \cite{Norte2016}. Similarly, only slight enhancements are observed for stoichiometric Si$_3$N$_4$ samples (not shown here).
	
There are still benefits of clamp-widening compared to regular strings in the diagonal orientation. As was shown in Figures~\ref{fig:3}a\&b, the $Q$ remains relatively stable over a range of $R$-values. This translates into an increased frequency, while keeping the $Q$ constant, thus increasing the $Q \times f$ product. For the sample presented in Figure \ref{fig:3}a, the $Q \times f$ product is increased slightly by a factor of 1.5 for $R = 50$ $\mu$m compared to $R = 0$ $\mu$m. Even larger enhancements can possibly be achieved by optimization of $R$ and $\sigma_p$. For force sensing applications, increasing both $Q$ and $f$ reduces the force noise, while increasing $Q$ increases the force responsivity.
	
The remainder of this section will focus on giving alternative explanations for the $Q$s of trampoline resonators. Norte et al. observed increasing $Q$s for decreasing Si$_3$N$_4$ thickness \cite{Norte2016}. This behavior is in contrast to what is expected for surface-loss limited $Q$s, where the decrease in $Q_{\text{int}}$ counteracts the enhancement from dissipation dilution. An increase in $Q$ for decreasing thickness would suggest that surface loss has (at least partially) been overcome, possibly a result of cleaning steps during fabrication. Alternatively, the reduced stress at the clamping could result in reduced radiation losses due to the larger acoustic mismatch \cite{Wilson-Rae2011}. The clamping region would then act as a phononic shield for the uniform stress region of the string and enhance the $Q$s.
	
\section{\label{sec:conclusion}Conclusion\protect}
The effect of clamp-widening on the quality factor of strained silicon nitride string resonators has been investigated for strings oriented both diagonally and perpendicularly with respect to the silicon frame. While $Q$ is only found to deteriorate with increasing clamp radius for perpendicularly oriented strings, a slight increase is observed for diagonally oriented strings at small radii. By studying the mode-dependence of the $Q$, it is found that the enhancement only occurs for the fundamental mode, while the $Q$ deteriorates for higher order modes compared to strings without widening. The results agree well with FEM simulations for both configurations. As such, large $Q$s observed for trampoline resonators cannot be explained through clamp-widening within the framework of dissipation dilution \cite{Norte2016}.
	
Nonetheless, clamp-widening does slightly increase the $Q \times f$ product, important for quantum optomechanics experiments. For force sensing applications, clamp-widening reduces force noise, while also increasing force responsivity. The method thus provides a simple means of improving various quantities. However, for fundamental applications, the method pales in comparison to the effectiveness of soft-clamping and strain engineering \cite{Tsaturyan2017,Ghadimi2017,Ghadimi2018}. Given that the data shown here does not reproduce the observed $Q$s using trampoline resonators, the effect is most likely intrinsic and a result of reduced surface losses. Negating surface losses can therefore be considered the last step towards reaching the ultimate limits of $Q$s in Si$_3$N$_4$ resonators and pave the way for performing quantum experiments at room-temperature.
	
\begin{acknowledgments}
	We wish to acknowledge the support from Sophia Ewert, Patrick Meyer, Johannes Schalko, and Niklas Luhmann for cleanroom fabrication.
\end{acknowledgments}
	
\nocite{*}
\bibliography{refs}% Produces the bibliography via BibTeX.

%merlin.mbs aipnum4-1.bst 2010-07-25 4.21a (PWD, AO, DPC) hacked
%Control: key (0)
%Control: author (8) initials jnrlst
%Control: editor formatted (1) identically to author
%Control: production of article title (0) allowed
%Control: page (1) range
%Control: year (1) truncated
%Control: production of eprint (0) enabled
\providecommand{\noopsort}[1]{}\providecommand{\singleletter}[1]{#1}%
\begin{thebibliography}{40}%
\makeatletter
\providecommand \@ifxundefined [1]{%
 \@ifx{#1\undefined}
}%
\providecommand \@ifnum [1]{%
 \ifnum #1\expandafter \@firstoftwo
 \else \expandafter \@secondoftwo
 \fi
}%
\providecommand \@ifx [1]{%
 \ifx #1\expandafter \@firstoftwo
 \else \expandafter \@secondoftwo
 \fi
}%
\providecommand \natexlab [1]{#1}%
\providecommand \enquote  [1]{``#1''}%
\providecommand \bibnamefont  [1]{#1}%
\providecommand \bibfnamefont [1]{#1}%
\providecommand \citenamefont [1]{#1}%
\providecommand \href@noop [0]{\@secondoftwo}%
\providecommand \href [0]{\begingroup \@sanitize@url \@href}%
\providecommand \@href[1]{\@@startlink{#1}\@@href}%
\providecommand \@@href[1]{\endgroup#1\@@endlink}%
\providecommand \@sanitize@url [0]{\catcode `\\12\catcode `\$12\catcode
  `\&12\catcode `\#12\catcode `\^12\catcode `\_12\catcode `\%12\relax}%
\providecommand \@@startlink[1]{}%
\providecommand \@@endlink[0]{}%
\providecommand \url  [0]{\begingroup\@sanitize@url \@url }%
\providecommand \@url [1]{\endgroup\@href {#1}{\urlprefix }}%
\providecommand \urlprefix  [0]{URL }%
\providecommand \Eprint [0]{\href }%
\providecommand \doibase [0]{http://dx.doi.org/}%
\providecommand \selectlanguage [0]{\@gobble}%
\providecommand \bibinfo  [0]{\@secondoftwo}%
\providecommand \bibfield  [0]{\@secondoftwo}%
\providecommand \translation [1]{[#1]}%
\providecommand \BibitemOpen [0]{}%
\providecommand \bibitemStop [0]{}%
\providecommand \bibitemNoStop [0]{.\EOS\space}%
\providecommand \EOS [0]{\spacefactor3000\relax}%
\providecommand \BibitemShut  [1]{\csname bibitem#1\endcsname}%
\let\auto@bib@innerbib\@empty
%</preamble>
\bibitem [{\citenamefont {Ekinci}\ and\ \citenamefont
  {Roukes}(2005)}]{Ekinci2005}%
  \BibitemOpen
  \bibfield  {author} {\bibinfo {author} {\bibfnamefont {K.~L.}\ \bibnamefont
  {Ekinci}}\ and\ \bibinfo {author} {\bibfnamefont {M.~L.}\ \bibnamefont
  {Roukes}},\ }\bibfield  {title} {\enquote {\bibinfo {title}
  {Nanoelectromechanical systems},}\ }\href {\doibase 10.1063/1.1927327}
  {\bibfield  {journal} {\bibinfo  {journal} {Review of Scientific
  Instruments}\ }\textbf {\bibinfo {volume} {76}},\ \bibinfo {pages} {061101}
  (\bibinfo {year} {2005})},\ \Eprint
  {http://arxiv.org/abs/https://doi.org/10.1063/1.1927327}
  {https://doi.org/10.1063/1.1927327} \BibitemShut {NoStop}%
\bibitem [{\citenamefont {Cleland}\ and\ \citenamefont
  {Roukes}(2002)}]{Cleland2002}%
  \BibitemOpen
  \bibfield  {author} {\bibinfo {author} {\bibfnamefont {A.~N.}\ \bibnamefont
  {Cleland}}\ and\ \bibinfo {author} {\bibfnamefont {M.~L.}\ \bibnamefont
  {Roukes}},\ }\bibfield  {title} {\enquote {\bibinfo {title} {Noise processes
  in nanomechanical resonators},}\ }\href {\doibase 10.1063/1.1499745}
  {\bibfield  {journal} {\bibinfo  {journal} {Journal of Applied Physics}\
  }\textbf {\bibinfo {volume} {92}},\ \bibinfo {pages} {2758--2769} (\bibinfo
  {year} {2002})},\ \Eprint
  {http://arxiv.org/abs/https://doi.org/10.1063/1.1499745}
  {https://doi.org/10.1063/1.1499745} \BibitemShut {NoStop}%
\bibitem [{\citenamefont {Stowe}\ \emph {et~al.}(1997)\citenamefont {Stowe},
  \citenamefont {Yasumura}, \citenamefont {Kenny}, \citenamefont {Botkin},
  \citenamefont {Wago},\ and\ \citenamefont {Rugar}}]{Stowe1997}%
  \BibitemOpen
  \bibfield  {author} {\bibinfo {author} {\bibfnamefont {T.~D.}\ \bibnamefont
  {Stowe}}, \bibinfo {author} {\bibfnamefont {K.}~\bibnamefont {Yasumura}},
  \bibinfo {author} {\bibfnamefont {T.~W.}\ \bibnamefont {Kenny}}, \bibinfo
  {author} {\bibfnamefont {D.}~\bibnamefont {Botkin}}, \bibinfo {author}
  {\bibfnamefont {K.}~\bibnamefont {Wago}}, \ and\ \bibinfo {author}
  {\bibfnamefont {D.}~\bibnamefont {Rugar}},\ }\bibfield  {title} {\enquote
  {\bibinfo {title} {Attonewton force detection using ultrathin silicon
  cantilevers},}\ }\href {\doibase 10.1063/1.119522} {\bibfield  {journal}
  {\bibinfo  {journal} {Applied Physics Letters}\ }\textbf {\bibinfo {volume}
  {71}},\ \bibinfo {pages} {288--290} (\bibinfo {year} {1997})},\ \Eprint
  {http://arxiv.org/abs/https://doi.org/10.1063/1.119522}
  {https://doi.org/10.1063/1.119522} \BibitemShut {NoStop}%
\bibitem [{\citenamefont {Rugar}\ \emph {et~al.}(2004)\citenamefont {Rugar},
  \citenamefont {Budakian}, \citenamefont {Mamin},\ and\ \citenamefont
  {Chui}}]{Rugar2004}%
  \BibitemOpen
  \bibfield  {author} {\bibinfo {author} {\bibfnamefont {D.}~\bibnamefont
  {Rugar}}, \bibinfo {author} {\bibfnamefont {R.}~\bibnamefont {Budakian}},
  \bibinfo {author} {\bibfnamefont {H.~J.}\ \bibnamefont {Mamin}}, \ and\
  \bibinfo {author} {\bibfnamefont {B.~W.}\ \bibnamefont {Chui}},\ }\bibfield
  {title} {\enquote {\bibinfo {title} {Single spin detection by magnetic
  resonance force microscopy},}\ }\href {\doibase 10.1038/nature02658}
  {\bibfield  {journal} {\bibinfo  {journal} {Nature}\ }\textbf {\bibinfo
  {volume} {430}},\ \bibinfo {pages} {329--332} (\bibinfo {year} {2004})},\
  \Eprint {http://arxiv.org/abs/https://doi.org/10.1038/nature02658}
  {https://doi.org/10.1038/nature02658} \BibitemShut {NoStop}%
\bibitem [{\citenamefont {Schmid}, \citenamefont {Villanueva},\ and\
  \citenamefont {Roukes}(2016)}]{Schmid2016}%
  \BibitemOpen
  \bibfield  {author} {\bibinfo {author} {\bibfnamefont {S.}~\bibnamefont
  {Schmid}}, \bibinfo {author} {\bibfnamefont {L.~G.}\ \bibnamefont
  {Villanueva}}, \ and\ \bibinfo {author} {\bibfnamefont {M.~L.}\ \bibnamefont
  {Roukes}},\ }\href@noop {} {\emph {\bibinfo {title} {Fundamentals of
  nanomechanical resonators}}}\ (\bibinfo  {publisher} {Springer International
  Edition},\ \bibinfo {year} {2016})\BibitemShut {NoStop}%
\bibitem [{\citenamefont {Aspelmeyer}, \citenamefont {Kippenberg},\ and\
  \citenamefont {Marquardt}(2014)}]{Aspelmeyer2014}%
  \BibitemOpen
  \bibfield  {author} {\bibinfo {author} {\bibfnamefont {M.}~\bibnamefont
  {Aspelmeyer}}, \bibinfo {author} {\bibfnamefont {T.~J.}\ \bibnamefont
  {Kippenberg}}, \ and\ \bibinfo {author} {\bibfnamefont {F.}~\bibnamefont
  {Marquardt}},\ }\bibfield  {title} {\enquote {\bibinfo {title} {Cavity
  optomechanics},}\ }\href {\doibase 10.1103/RevModPhys.86.1391} {\bibfield
  {journal} {\bibinfo  {journal} {Rev. Mod. Phys.}\ }\textbf {\bibinfo {volume}
  {86}},\ \bibinfo {pages} {1391--1452} (\bibinfo {year} {2014})}\BibitemShut
  {NoStop}%
\bibitem [{\citenamefont {Carr}\ \emph {et~al.}(1999)\citenamefont {Carr},
  \citenamefont {Evoy}, \citenamefont {Sekaric}, \citenamefont {Craighead},\
  and\ \citenamefont {Parpia}}]{Carr1999}%
  \BibitemOpen
  \bibfield  {author} {\bibinfo {author} {\bibfnamefont {D.~W.}\ \bibnamefont
  {Carr}}, \bibinfo {author} {\bibfnamefont {S.}~\bibnamefont {Evoy}}, \bibinfo
  {author} {\bibfnamefont {L.}~\bibnamefont {Sekaric}}, \bibinfo {author}
  {\bibfnamefont {H.~G.}\ \bibnamefont {Craighead}}, \ and\ \bibinfo {author}
  {\bibfnamefont {J.~M.}\ \bibnamefont {Parpia}},\ }\bibfield  {title}
  {\enquote {\bibinfo {title} {Measurement of mechanical resonance and losses
  in nanometer scale silicon wires},}\ }\href {\doibase 10.1063/1.124554}
  {\bibfield  {journal} {\bibinfo  {journal} {Applied Physics Letters}\
  }\textbf {\bibinfo {volume} {75}},\ \bibinfo {pages} {920--922} (\bibinfo
  {year} {1999})},\ \Eprint
  {http://arxiv.org/abs/https://doi.org/10.1063/1.124554}
  {https://doi.org/10.1063/1.124554} \BibitemShut {NoStop}%
\bibitem [{\citenamefont {Verbridge}\ \emph {et~al.}(2006)\citenamefont
  {Verbridge}, \citenamefont {Parpia}, \citenamefont {Reichenbach},
  \citenamefont {Bellan},\ and\ \citenamefont {Craighead}}]{Verbridge2006}%
  \BibitemOpen
  \bibfield  {author} {\bibinfo {author} {\bibfnamefont {S.~S.}\ \bibnamefont
  {Verbridge}}, \bibinfo {author} {\bibfnamefont {J.~M.}\ \bibnamefont
  {Parpia}}, \bibinfo {author} {\bibfnamefont {R.~B.}\ \bibnamefont
  {Reichenbach}}, \bibinfo {author} {\bibfnamefont {L.~M.}\ \bibnamefont
  {Bellan}}, \ and\ \bibinfo {author} {\bibfnamefont {H.~G.}\ \bibnamefont
  {Craighead}},\ }\bibfield  {title} {\enquote {\bibinfo {title} {High quality
  factor resonance at room temperature with nanostrings under high tensile
  stress},}\ }\href {\doibase 10.1063/1.2204829} {\bibfield  {journal}
  {\bibinfo  {journal} {Journal of Applied Physics}\ }\textbf {\bibinfo
  {volume} {99}},\ \bibinfo {pages} {124304} (\bibinfo {year} {2006})},\
  \Eprint {http://arxiv.org/abs/https://doi.org/10.1063/1.2204829}
  {https://doi.org/10.1063/1.2204829} \BibitemShut {NoStop}%
\bibitem [{\citenamefont {Verbridge}, \citenamefont {Craighead},\ and\
  \citenamefont {Parpia}(2008)}]{Verbridge2008}%
  \BibitemOpen
  \bibfield  {author} {\bibinfo {author} {\bibfnamefont {S.~S.}\ \bibnamefont
  {Verbridge}}, \bibinfo {author} {\bibfnamefont {H.~G.}\ \bibnamefont
  {Craighead}}, \ and\ \bibinfo {author} {\bibfnamefont {J.~M.}\ \bibnamefont
  {Parpia}},\ }\bibfield  {title} {\enquote {\bibinfo {title} {A megahertz
  nanomechanical resonator with room temperature quality factor over a
  million},}\ }\href {\doibase 10.1063/1.2822406} {\bibfield  {journal}
  {\bibinfo  {journal} {Applied Physics Letters}\ }\textbf {\bibinfo {volume}
  {92}},\ \bibinfo {pages} {013112} (\bibinfo {year} {2008})},\ \Eprint
  {http://arxiv.org/abs/https://aip.scitation.org/doi/pdf/10.1063/1.2822406}
  {https://aip.scitation.org/doi/pdf/10.1063/1.2822406} \BibitemShut {NoStop}%
\bibitem [{\citenamefont {Verbridge}\ \emph {et~al.}(2007)\citenamefont
  {Verbridge}, \citenamefont {Shapiro}, \citenamefont {Craighead},\ and\
  \citenamefont {Parpia}}]{Verbridge2007}%
  \BibitemOpen
  \bibfield  {author} {\bibinfo {author} {\bibfnamefont {S.~S.}\ \bibnamefont
  {Verbridge}}, \bibinfo {author} {\bibfnamefont {D.~F.}\ \bibnamefont
  {Shapiro}}, \bibinfo {author} {\bibfnamefont {H.~G.}\ \bibnamefont
  {Craighead}}, \ and\ \bibinfo {author} {\bibfnamefont {J.~M.}\ \bibnamefont
  {Parpia}},\ }\bibfield  {title} {\enquote {\bibinfo {title} {Macroscopic
  tuning of nanomechanics: Substrate bending for reversible control of
  frequency and quality factor of nanostring resonators},}\ }\href {\doibase
  10.1021/nl070716t} {\bibfield  {journal} {\bibinfo  {journal} {Nano Letters}\
  }\textbf {\bibinfo {volume} {7}},\ \bibinfo {pages} {1728--1735} (\bibinfo
  {year} {2007})},\ \bibinfo {note} {pMID: 17497822},\ \Eprint
  {http://arxiv.org/abs/https://doi.org/10.1021/nl070716t}
  {https://doi.org/10.1021/nl070716t} \BibitemShut {NoStop}%
\bibitem [{\citenamefont {González}\ and\ \citenamefont
  {Saulson}(1994)}]{Gonzalez1994}%
  \BibitemOpen
  \bibfield  {author} {\bibinfo {author} {\bibfnamefont {G.~I.}\ \bibnamefont
  {González}}\ and\ \bibinfo {author} {\bibfnamefont {P.~R.}\ \bibnamefont
  {Saulson}},\ }\bibfield  {title} {\enquote {\bibinfo {title} {Brownian motion
  of a mass suspended by an anelastic wire},}\ }\href {\doibase
  10.1121/1.410467} {\bibfield  {journal} {\bibinfo  {journal} {The Journal of
  the Acoustical Society of America}\ }\textbf {\bibinfo {volume} {96}},\
  \bibinfo {pages} {207--212} (\bibinfo {year} {1994})},\ \Eprint
  {http://arxiv.org/abs/https://doi.org/10.1121/1.410467}
  {https://doi.org/10.1121/1.410467} \BibitemShut {NoStop}%
\bibitem [{\citenamefont {Huang}\ and\ \citenamefont
  {Saulson}(1998)}]{Huang1998}%
  \BibitemOpen
  \bibfield  {author} {\bibinfo {author} {\bibfnamefont {Y.~L.}\ \bibnamefont
  {Huang}}\ and\ \bibinfo {author} {\bibfnamefont {P.~R.}\ \bibnamefont
  {Saulson}},\ }\bibfield  {title} {\enquote {\bibinfo {title} {Dissipation
  mechanisms in pendulums and their implications for gravitational wave
  interferometers},}\ }\href {\doibase 10.1063/1.1148692} {\bibfield  {journal}
  {\bibinfo  {journal} {Review of Scientific Instruments}\ }\textbf {\bibinfo
  {volume} {69}},\ \bibinfo {pages} {544--553} (\bibinfo {year} {1998})},\
  \Eprint {http://arxiv.org/abs/https://doi.org/10.1063/1.1148692}
  {https://doi.org/10.1063/1.1148692} \BibitemShut {NoStop}%
\bibitem [{\citenamefont {Unterreithmeier}, \citenamefont {Faust},\ and\
  \citenamefont {Kotthaus}(2010)}]{Unterreithmeier2010}%
  \BibitemOpen
  \bibfield  {author} {\bibinfo {author} {\bibfnamefont {Q.~P.}\ \bibnamefont
  {Unterreithmeier}}, \bibinfo {author} {\bibfnamefont {T.}~\bibnamefont
  {Faust}}, \ and\ \bibinfo {author} {\bibfnamefont {J.~P.}\ \bibnamefont
  {Kotthaus}},\ }\bibfield  {title} {\enquote {\bibinfo {title} {Damping of
  nanomechanical resonators},}\ }\href {\doibase
  10.1103/PhysRevLett.105.027205} {\bibfield  {journal} {\bibinfo  {journal}
  {Phys. Rev. Lett.}\ }\textbf {\bibinfo {volume} {105}},\ \bibinfo {pages}
  {027205} (\bibinfo {year} {2010})}\BibitemShut {NoStop}%
\bibitem [{\citenamefont {Schmid}\ \emph {et~al.}(2011)\citenamefont {Schmid},
  \citenamefont {Jensen}, \citenamefont {Nielsen},\ and\ \citenamefont
  {Boisen}}]{Schmid2011}%
  \BibitemOpen
  \bibfield  {author} {\bibinfo {author} {\bibfnamefont {S.}~\bibnamefont
  {Schmid}}, \bibinfo {author} {\bibfnamefont {K.~D.}\ \bibnamefont {Jensen}},
  \bibinfo {author} {\bibfnamefont {K.~H.}\ \bibnamefont {Nielsen}}, \ and\
  \bibinfo {author} {\bibfnamefont {A.}~\bibnamefont {Boisen}},\ }\bibfield
  {title} {\enquote {\bibinfo {title} {Damping mechanisms in high-$q$ micro and
  nanomechanical string resonators},}\ }\href {\doibase
  10.1103/PhysRevB.84.165307} {\bibfield  {journal} {\bibinfo  {journal} {Phys.
  Rev. B}\ }\textbf {\bibinfo {volume} {84}},\ \bibinfo {pages} {165307}
  (\bibinfo {year} {2011})}\BibitemShut {NoStop}%
\bibitem [{\citenamefont {Schmid}\ and\ \citenamefont
  {Hierold}(2008)}]{Schmid2008}%
  \BibitemOpen
  \bibfield  {author} {\bibinfo {author} {\bibfnamefont {S.}~\bibnamefont
  {Schmid}}\ and\ \bibinfo {author} {\bibfnamefont {C.}~\bibnamefont
  {Hierold}},\ }\bibfield  {title} {\enquote {\bibinfo {title} {Damping
  mechanisms of single-clamped and prestressed double-clamped resonant polymer
  microbeams},}\ }\href {\doibase 10.1063/1.3008032} {\bibfield  {journal}
  {\bibinfo  {journal} {Journal of Applied Physics}\ }\textbf {\bibinfo
  {volume} {104}},\ \bibinfo {pages} {093516} (\bibinfo {year} {2008})},\
  \Eprint {http://arxiv.org/abs/https://doi.org/10.1063/1.3008032}
  {https://doi.org/10.1063/1.3008032} \BibitemShut {NoStop}%
\bibitem [{\citenamefont {Cole}\ \emph {et~al.}(2014)\citenamefont {Cole},
  \citenamefont {Yu}, \citenamefont {Gärtner}, \citenamefont {Siquans},
  \citenamefont {Moghadas~Nia}, \citenamefont {Schmöle}, \citenamefont
  {Hoelscher-Obermaier}, \citenamefont {Purdy}, \citenamefont {Wieczorek},
  \citenamefont {Regal},\ and\ \citenamefont {Aspelmeyer}}]{Cole2014}%
  \BibitemOpen
  \bibfield  {author} {\bibinfo {author} {\bibfnamefont {G.~D.}\ \bibnamefont
  {Cole}}, \bibinfo {author} {\bibfnamefont {P.-L.}\ \bibnamefont {Yu}},
  \bibinfo {author} {\bibfnamefont {C.}~\bibnamefont {Gärtner}}, \bibinfo
  {author} {\bibfnamefont {K.}~\bibnamefont {Siquans}}, \bibinfo {author}
  {\bibfnamefont {R.}~\bibnamefont {Moghadas~Nia}}, \bibinfo {author}
  {\bibfnamefont {J.}~\bibnamefont {Schmöle}}, \bibinfo {author}
  {\bibfnamefont {J.}~\bibnamefont {Hoelscher-Obermaier}}, \bibinfo {author}
  {\bibfnamefont {T.~P.}\ \bibnamefont {Purdy}}, \bibinfo {author}
  {\bibfnamefont {W.}~\bibnamefont {Wieczorek}}, \bibinfo {author}
  {\bibfnamefont {C.~A.}\ \bibnamefont {Regal}}, \ and\ \bibinfo {author}
  {\bibfnamefont {M.}~\bibnamefont {Aspelmeyer}},\ }\bibfield  {title}
  {\enquote {\bibinfo {title} {Tensile-strained inxga1-xp membranes for cavity
  optomechanics},}\ }\href {\doibase 10.1063/1.4879755} {\bibfield  {journal}
  {\bibinfo  {journal} {Applied Physics Letters}\ }\textbf {\bibinfo {volume}
  {104}},\ \bibinfo {pages} {201908} (\bibinfo {year} {2014})},\ \Eprint
  {http://arxiv.org/abs/https://doi.org/10.1063/1.4879755}
  {https://doi.org/10.1063/1.4879755} \BibitemShut {NoStop}%
\bibitem [{\citenamefont {Cagnoli}\ \emph {et~al.}(2000)\citenamefont
  {Cagnoli}, \citenamefont {Hough}, \citenamefont {DeBra}, \citenamefont
  {Fejer}, \citenamefont {Gustafson}, \citenamefont {Rowan},\ and\
  \citenamefont {Mitrofanov}}]{Cagnoli2000}%
  \BibitemOpen
  \bibfield  {author} {\bibinfo {author} {\bibfnamefont {G.}~\bibnamefont
  {Cagnoli}}, \bibinfo {author} {\bibfnamefont {J.}~\bibnamefont {Hough}},
  \bibinfo {author} {\bibfnamefont {D.}~\bibnamefont {DeBra}}, \bibinfo
  {author} {\bibfnamefont {M.}~\bibnamefont {Fejer}}, \bibinfo {author}
  {\bibfnamefont {E.}~\bibnamefont {Gustafson}}, \bibinfo {author}
  {\bibfnamefont {S.}~\bibnamefont {Rowan}}, \ and\ \bibinfo {author}
  {\bibfnamefont {V.}~\bibnamefont {Mitrofanov}},\ }\bibfield  {title}
  {\enquote {\bibinfo {title} {Damping dilution factor for a pendulum in an
  interferometric gravitational waves detector},}\ }\href {\doibase
  https://doi.org/10.1016/S0375-9601(00)00411-4} {\bibfield  {journal}
  {\bibinfo  {journal} {Physics Letters A}\ }\textbf {\bibinfo {volume}
  {272}},\ \bibinfo {pages} {39 -- 45} (\bibinfo {year} {2000})}\BibitemShut
  {NoStop}%
\bibitem [{\citenamefont {Fedorov}\ \emph {et~al.}(2019)\citenamefont
  {Fedorov}, \citenamefont {Engelsen}, \citenamefont {Ghadimi}, \citenamefont
  {Bereyhi}, \citenamefont {Schilling}, \citenamefont {Wilson},\ and\
  \citenamefont {Kippenberg}}]{Fedorov2019}%
  \BibitemOpen
  \bibfield  {author} {\bibinfo {author} {\bibfnamefont {S.~A.}\ \bibnamefont
  {Fedorov}}, \bibinfo {author} {\bibfnamefont {N.~J.}\ \bibnamefont
  {Engelsen}}, \bibinfo {author} {\bibfnamefont {A.~H.}\ \bibnamefont
  {Ghadimi}}, \bibinfo {author} {\bibfnamefont {M.~J.}\ \bibnamefont
  {Bereyhi}}, \bibinfo {author} {\bibfnamefont {R.}~\bibnamefont {Schilling}},
  \bibinfo {author} {\bibfnamefont {D.~J.}\ \bibnamefont {Wilson}}, \ and\
  \bibinfo {author} {\bibfnamefont {T.~J.}\ \bibnamefont {Kippenberg}},\
  }\bibfield  {title} {\enquote {\bibinfo {title} {Generalized dissipation
  dilution in strained mechanical resonators},}\ }\href {\doibase
  10.1103/PhysRevB.99.054107} {\bibfield  {journal} {\bibinfo  {journal} {Phys.
  Rev. B}\ }\textbf {\bibinfo {volume} {99}},\ \bibinfo {pages} {054107}
  (\bibinfo {year} {2019})}\BibitemShut {NoStop}%
\bibitem [{\citenamefont {Yu}\ \emph {et~al.}(2014)\citenamefont {Yu},
  \citenamefont {Cicak}, \citenamefont {Kampel}, \citenamefont {Tsaturyan},
  \citenamefont {Purdy}, \citenamefont {Simmonds},\ and\ \citenamefont
  {Regal}}]{Yu2014}%
  \BibitemOpen
  \bibfield  {author} {\bibinfo {author} {\bibfnamefont {P.-L.}\ \bibnamefont
  {Yu}}, \bibinfo {author} {\bibfnamefont {K.}~\bibnamefont {Cicak}}, \bibinfo
  {author} {\bibfnamefont {N.~S.}\ \bibnamefont {Kampel}}, \bibinfo {author}
  {\bibfnamefont {Y.}~\bibnamefont {Tsaturyan}}, \bibinfo {author}
  {\bibfnamefont {T.~P.}\ \bibnamefont {Purdy}}, \bibinfo {author}
  {\bibfnamefont {R.~W.}\ \bibnamefont {Simmonds}}, \ and\ \bibinfo {author}
  {\bibfnamefont {C.~A.}\ \bibnamefont {Regal}},\ }\bibfield  {title} {\enquote
  {\bibinfo {title} {A phononic bandgap shield for high-q membrane
  microresonators},}\ }\href {\doibase 10.1063/1.4862031} {\bibfield  {journal}
  {\bibinfo  {journal} {Applied Physics Letters}\ }\textbf {\bibinfo {volume}
  {104}},\ \bibinfo {pages} {023510} (\bibinfo {year} {2014})},\ \Eprint
  {http://arxiv.org/abs/https://doi.org/10.1063/1.4862031}
  {https://doi.org/10.1063/1.4862031} \BibitemShut {NoStop}%
\bibitem [{\citenamefont {Tsaturyan}\ \emph {et~al.}(2014)\citenamefont
  {Tsaturyan}, \citenamefont {Barg}, \citenamefont {Simonsen}, \citenamefont
  {Villanueva}, \citenamefont {Schmid}, \citenamefont {Schliesser},\ and\
  \citenamefont {Polzik}}]{Tsaturyan2014}%
  \BibitemOpen
  \bibfield  {author} {\bibinfo {author} {\bibfnamefont {Y.}~\bibnamefont
  {Tsaturyan}}, \bibinfo {author} {\bibfnamefont {A.}~\bibnamefont {Barg}},
  \bibinfo {author} {\bibfnamefont {A.}~\bibnamefont {Simonsen}}, \bibinfo
  {author} {\bibfnamefont {L.~G.}\ \bibnamefont {Villanueva}}, \bibinfo
  {author} {\bibfnamefont {S.}~\bibnamefont {Schmid}}, \bibinfo {author}
  {\bibfnamefont {A.}~\bibnamefont {Schliesser}}, \ and\ \bibinfo {author}
  {\bibfnamefont {E.~S.}\ \bibnamefont {Polzik}},\ }\bibfield  {title}
  {\enquote {\bibinfo {title} {Demonstration of suppressed phonon tunneling
  losses in phononic bandgap shielded membrane resonators for high-q
  optomechanics},}\ }\href {\doibase 10.1364/OE.22.006810} {\bibfield
  {journal} {\bibinfo  {journal} {Opt. Express}\ }\textbf {\bibinfo {volume}
  {22}},\ \bibinfo {pages} {6810--6821} (\bibinfo {year} {2014})}\BibitemShut
  {NoStop}%
\bibitem [{\citenamefont {Tsaturyan}\ \emph {et~al.}(2017)\citenamefont
  {Tsaturyan}, \citenamefont {Barg}, \citenamefont {Polzik},\ and\
  \citenamefont {Schliesser}}]{Tsaturyan2017}%
  \BibitemOpen
  \bibfield  {author} {\bibinfo {author} {\bibfnamefont {Y.}~\bibnamefont
  {Tsaturyan}}, \bibinfo {author} {\bibfnamefont {A.}~\bibnamefont {Barg}},
  \bibinfo {author} {\bibfnamefont {E.~S.}\ \bibnamefont {Polzik}}, \ and\
  \bibinfo {author} {\bibfnamefont {A.}~\bibnamefont {Schliesser}},\ }\bibfield
   {title} {\enquote {\bibinfo {title} {Ultracoherent nanomechanical resonators
  via soft clamping and dissipation dilution},}\ }\href {\doibase
  10.1038/nnano.2017.101} {\bibfield  {journal} {\bibinfo  {journal} {Nature
  Nanotechnology}\ }\textbf {\bibinfo {volume} {12}},\ \bibinfo {pages}
  {776--783} (\bibinfo {year} {2017})},\ \Eprint
  {http://arxiv.org/abs/https://doi.org/10.1038/nnano.2017.101}
  {https://doi.org/10.1038/nnano.2017.101} \BibitemShut {NoStop}%
\bibitem [{\citenamefont {Ghadimi}, \citenamefont {Wilson},\ and\ \citenamefont
  {Kippenberg}(2017)}]{Ghadimi2017}%
  \BibitemOpen
  \bibfield  {author} {\bibinfo {author} {\bibfnamefont {A.~H.}\ \bibnamefont
  {Ghadimi}}, \bibinfo {author} {\bibfnamefont {D.~J.}\ \bibnamefont {Wilson}},
  \ and\ \bibinfo {author} {\bibfnamefont {T.~J.}\ \bibnamefont {Kippenberg}},\
  }\bibfield  {title} {\enquote {\bibinfo {title} {Radiation and internal loss
  engineering of high-stress silicon nitride nanobeams},}\ }\href {\doibase
  10.1021/acs.nanolett.7b00573} {\bibfield  {journal} {\bibinfo  {journal}
  {Nano Letters}\ }\textbf {\bibinfo {volume} {17}},\ \bibinfo {pages}
  {3501--3505} (\bibinfo {year} {2017})},\ \bibinfo {note} {pMID: 28362505},\
  \Eprint {http://arxiv.org/abs/https://doi.org/10.1021/acs.nanolett.7b00573}
  {https://doi.org/10.1021/acs.nanolett.7b00573} \BibitemShut {NoStop}%
\bibitem [{\citenamefont {Ghadimi}\ \emph {et~al.}(2018)\citenamefont
  {Ghadimi}, \citenamefont {Fedorov}, \citenamefont {Engelsen}, \citenamefont
  {Bereyhi}, \citenamefont {Schilling}, \citenamefont {Wilson},\ and\
  \citenamefont {Kippenberg}}]{Ghadimi2018}%
  \BibitemOpen
  \bibfield  {author} {\bibinfo {author} {\bibfnamefont {A.~H.}\ \bibnamefont
  {Ghadimi}}, \bibinfo {author} {\bibfnamefont {S.~A.}\ \bibnamefont
  {Fedorov}}, \bibinfo {author} {\bibfnamefont {N.~J.}\ \bibnamefont
  {Engelsen}}, \bibinfo {author} {\bibfnamefont {M.~J.}\ \bibnamefont
  {Bereyhi}}, \bibinfo {author} {\bibfnamefont {R.}~\bibnamefont {Schilling}},
  \bibinfo {author} {\bibfnamefont {D.~J.}\ \bibnamefont {Wilson}}, \ and\
  \bibinfo {author} {\bibfnamefont {T.~J.}\ \bibnamefont {Kippenberg}},\
  }\bibfield  {title} {\enquote {\bibinfo {title} {Elastic strain engineering
  for ultralow mechanical dissipation},}\ }\href {\doibase
  10.1126/science.aar6939} {\bibfield  {journal} {\bibinfo  {journal}
  {Science}\ }\textbf {\bibinfo {volume} {360}},\ \bibinfo {pages} {764--768}
  (\bibinfo {year} {2018})},\ \Eprint
  {http://arxiv.org/abs/http://science.sciencemag.org/content/360/6390/764.full.pdf}
  {http://science.sciencemag.org/content/360/6390/764.full.pdf} \BibitemShut
  {NoStop}%
\bibitem [{\citenamefont {Villanueva}\ and\ \citenamefont
  {Schmid}(2014)}]{Villanueva2014}%
  \BibitemOpen
  \bibfield  {author} {\bibinfo {author} {\bibfnamefont {L.~G.}\ \bibnamefont
  {Villanueva}}\ and\ \bibinfo {author} {\bibfnamefont {S.}~\bibnamefont
  {Schmid}},\ }\bibfield  {title} {\enquote {\bibinfo {title} {Evidence of
  surface loss as ubiquitous limiting damping mechanism in sin micro- and
  nanomechanical resonators},}\ }\href {\doibase
  10.1103/PhysRevLett.113.227201} {\bibfield  {journal} {\bibinfo  {journal}
  {Phys. Rev. Lett.}\ }\textbf {\bibinfo {volume} {113}},\ \bibinfo {pages}
  {227201} (\bibinfo {year} {2014})}\BibitemShut {NoStop}%
\bibitem [{\citenamefont {Norte}, \citenamefont {Moura},\ and\ \citenamefont
  {Gr\"oblacher}(2016)}]{Norte2016}%
  \BibitemOpen
  \bibfield  {author} {\bibinfo {author} {\bibfnamefont {R.~A.}\ \bibnamefont
  {Norte}}, \bibinfo {author} {\bibfnamefont {J.~P.}\ \bibnamefont {Moura}}, \
  and\ \bibinfo {author} {\bibfnamefont {S.}~\bibnamefont {Gr\"oblacher}},\
  }\bibfield  {title} {\enquote {\bibinfo {title} {Mechanical resonators for
  quantum optomechanics experiments at room temperature},}\ }\href {\doibase
  10.1103/PhysRevLett.116.147202} {\bibfield  {journal} {\bibinfo  {journal}
  {Phys. Rev. Lett.}\ }\textbf {\bibinfo {volume} {116}},\ \bibinfo {pages}
  {147202} (\bibinfo {year} {2016})}\BibitemShut {NoStop}%
\bibitem [{\citenamefont {Reinhardt}\ \emph {et~al.}(2016)\citenamefont
  {Reinhardt}, \citenamefont {M\"uller}, \citenamefont {Bourassa},\ and\
  \citenamefont {Sankey}}]{Reinhardt2016}%
  \BibitemOpen
  \bibfield  {author} {\bibinfo {author} {\bibfnamefont {C.}~\bibnamefont
  {Reinhardt}}, \bibinfo {author} {\bibfnamefont {T.}~\bibnamefont {M\"uller}},
  \bibinfo {author} {\bibfnamefont {A.}~\bibnamefont {Bourassa}}, \ and\
  \bibinfo {author} {\bibfnamefont {J.~C.}\ \bibnamefont {Sankey}},\ }\bibfield
   {title} {\enquote {\bibinfo {title} {Ultralow-noise sin trampoline
  resonators for sensing and optomechanics},}\ }\href {\doibase
  10.1103/PhysRevX.6.021001} {\bibfield  {journal} {\bibinfo  {journal} {Phys.
  Rev. X}\ }\textbf {\bibinfo {volume} {6}},\ \bibinfo {pages} {021001}
  (\bibinfo {year} {2016})}\BibitemShut {NoStop}%
\bibitem [{\citenamefont {Fischer}\ \emph {et~al.}(2019)\citenamefont
  {Fischer}, \citenamefont {McNally}, \citenamefont {Reetz}, \citenamefont
  {Assumpcao}, \citenamefont {Knief}, \citenamefont {Lin},\ and\ \citenamefont
  {Regal}}]{Fischer2019}%
  \BibitemOpen
  \bibfield  {author} {\bibinfo {author} {\bibfnamefont {R.}~\bibnamefont
  {Fischer}}, \bibinfo {author} {\bibfnamefont {D.~P.}\ \bibnamefont
  {McNally}}, \bibinfo {author} {\bibfnamefont {C.}~\bibnamefont {Reetz}},
  \bibinfo {author} {\bibfnamefont {G.~G.~T.}\ \bibnamefont {Assumpcao}},
  \bibinfo {author} {\bibfnamefont {T.~R.}\ \bibnamefont {Knief}}, \bibinfo
  {author} {\bibfnamefont {Y.}~\bibnamefont {Lin}}, \ and\ \bibinfo {author}
  {\bibfnamefont {C.~A.}\ \bibnamefont {Regal}},\ }\bibfield  {title} {\enquote
  {\bibinfo {title} {Spin detection with a micromechanical trampoline: Towards
  magnetic resonance microscopy harnessing cavity optomechanics},}\ }\href
  {http://iopscience.iop.org/10.1088/1367-2630/ab117a} {\bibfield  {journal}
  {\bibinfo  {journal} {New Journal of Physics}\ } (\bibinfo {year}
  {2019})}\BibitemShut {NoStop}%
\bibitem [{\citenamefont {Bereyhi}\ \emph {et~al.}(2019)\citenamefont
  {Bereyhi}, \citenamefont {Beccari}, \citenamefont {Fedorov}, \citenamefont
  {Ghadimi}, \citenamefont {Schilling}, \citenamefont {Wilson}, \citenamefont
  {Engelsen},\ and\ \citenamefont {Kippenberg}}]{Bereyhi2019}%
  \BibitemOpen
  \bibfield  {author} {\bibinfo {author} {\bibfnamefont {M.~J.}\ \bibnamefont
  {Bereyhi}}, \bibinfo {author} {\bibfnamefont {A.}~\bibnamefont {Beccari}},
  \bibinfo {author} {\bibfnamefont {S.~A.}\ \bibnamefont {Fedorov}}, \bibinfo
  {author} {\bibfnamefont {A.~H.}\ \bibnamefont {Ghadimi}}, \bibinfo {author}
  {\bibfnamefont {R.}~\bibnamefont {Schilling}}, \bibinfo {author}
  {\bibfnamefont {D.~J.}\ \bibnamefont {Wilson}}, \bibinfo {author}
  {\bibfnamefont {N.~J.}\ \bibnamefont {Engelsen}}, \ and\ \bibinfo {author}
  {\bibfnamefont {T.~J.}\ \bibnamefont {Kippenberg}},\ }\bibfield  {title}
  {\enquote {\bibinfo {title} {Clamp-tapering increases the quality factor of
  stressed nanobeams},}\ }\href {\doibase 10.1021/acs.nanolett.8b04942}
  {\bibfield  {journal} {\bibinfo  {journal} {Nano Letters}\ }\textbf {\bibinfo
  {volume} {19}},\ \bibinfo {pages} {2329--2333} (\bibinfo {year} {2019})},\
  \bibinfo {note} {pMID: 30811943},\ \Eprint
  {http://arxiv.org/abs/https://doi.org/10.1021/acs.nanolett.8b04942}
  {https://doi.org/10.1021/acs.nanolett.8b04942} \BibitemShut {NoStop}%
\bibitem [{\citenamefont {Schmid}, \citenamefont {Dohn},\ and\ \citenamefont
  {Boisen}(2010)}]{Schmid2010}%
  \BibitemOpen
  \bibfield  {author} {\bibinfo {author} {\bibfnamefont {S.}~\bibnamefont
  {Schmid}}, \bibinfo {author} {\bibfnamefont {S.}~\bibnamefont {Dohn}}, \ and\
  \bibinfo {author} {\bibfnamefont {A.}~\bibnamefont {Boisen}},\ }\bibfield
  {title} {\enquote {\bibinfo {title} {Real-time particle mass spectrometry
  based on resonant micro strings},}\ }\href {\doibase 10.3390/s100908092}
  {\bibfield  {journal} {\bibinfo  {journal} {Sensors}\ }\textbf {\bibinfo
  {volume} {10}},\ \bibinfo {pages} {8092--8100} (\bibinfo {year}
  {2010})}\BibitemShut {NoStop}%
\bibitem [{\citenamefont {Luhmann}\ \emph {et~al.}(2017)\citenamefont
  {Luhmann}, \citenamefont {Jachimowicz}, \citenamefont {Schalko},
  \citenamefont {Sadeghi}, \citenamefont {Sauer}, \citenamefont
  {Foelske-Schmitz},\ and\ \citenamefont {Schmid}}]{Luhmann2017}%
  \BibitemOpen
  \bibfield  {author} {\bibinfo {author} {\bibfnamefont {N.}~\bibnamefont
  {Luhmann}}, \bibinfo {author} {\bibfnamefont {A.}~\bibnamefont
  {Jachimowicz}}, \bibinfo {author} {\bibfnamefont {J.}~\bibnamefont
  {Schalko}}, \bibinfo {author} {\bibfnamefont {P.}~\bibnamefont {Sadeghi}},
  \bibinfo {author} {\bibfnamefont {M.}~\bibnamefont {Sauer}}, \bibinfo
  {author} {\bibfnamefont {A.}~\bibnamefont {Foelske-Schmitz}}, \ and\ \bibinfo
  {author} {\bibfnamefont {S.}~\bibnamefont {Schmid}},\ }\bibfield  {title}
  {\enquote {\bibinfo {title} {Effect of oxygen plasma on nanomechanical
  silicon nitride resonators},}\ }\href {\doibase 10.1063/1.4989775} {\bibfield
   {journal} {\bibinfo  {journal} {Applied Physics Letters}\ }\textbf {\bibinfo
  {volume} {111}},\ \bibinfo {pages} {063103} (\bibinfo {year} {2017})},\
  \Eprint {http://arxiv.org/abs/https://doi.org/10.1063/1.4989775}
  {https://doi.org/10.1063/1.4989775} \BibitemShut {NoStop}%
\bibitem [{\citenamefont {{Schmid}}, \citenamefont {{Malm}},\ and\
  \citenamefont {{Boisen}}(2011)}]{Schmid2011_2}%
  \BibitemOpen
  \bibfield  {author} {\bibinfo {author} {\bibfnamefont {S.}~\bibnamefont
  {{Schmid}}}, \bibinfo {author} {\bibfnamefont {B.}~\bibnamefont {{Malm}}}, \
  and\ \bibinfo {author} {\bibfnamefont {A.}~\bibnamefont {{Boisen}}},\
  }\bibfield  {title} {\enquote {\bibinfo {title} {Quality factor improvement
  of silicon nitride micro string resonators},}\ }in\ \href {\doibase
  10.1109/MEMSYS.2011.5734466} {\emph {\bibinfo {booktitle} {2011 IEEE 24th
  International Conference on Micro Electro Mechanical Systems}}}\ (\bibinfo
  {year} {2011})\ pp.\ \bibinfo {pages} {481--484}\BibitemShut {NoStop}%
\bibitem [{\citenamefont {Yu}, \citenamefont {Purdy},\ and\ \citenamefont
  {Regal}(2012)}]{Yu2012}%
  \BibitemOpen
  \bibfield  {author} {\bibinfo {author} {\bibfnamefont {P.-L.}\ \bibnamefont
  {Yu}}, \bibinfo {author} {\bibfnamefont {T.~P.}\ \bibnamefont {Purdy}}, \
  and\ \bibinfo {author} {\bibfnamefont {C.~A.}\ \bibnamefont {Regal}},\
  }\bibfield  {title} {\enquote {\bibinfo {title} {Control of material damping
  in high-$q$ membrane microresonators},}\ }\href {\doibase
  10.1103/PhysRevLett.108.083603} {\bibfield  {journal} {\bibinfo  {journal}
  {Phys. Rev. Lett.}\ }\textbf {\bibinfo {volume} {108}},\ \bibinfo {pages}
  {083603} (\bibinfo {year} {2012})}\BibitemShut {NoStop}%
\bibitem [{\citenamefont {Wilson-Rae}\ \emph {et~al.}(2011)\citenamefont
  {Wilson-Rae}, \citenamefont {Barton}, \citenamefont {Verbridge},
  \citenamefont {Southworth}, \citenamefont {Ilic}, \citenamefont {Craighead},\
  and\ \citenamefont {Parpia}}]{Wilson-Rae2011}%
  \BibitemOpen
  \bibfield  {author} {\bibinfo {author} {\bibfnamefont {I.}~\bibnamefont
  {Wilson-Rae}}, \bibinfo {author} {\bibfnamefont {R.~A.}\ \bibnamefont
  {Barton}}, \bibinfo {author} {\bibfnamefont {S.~S.}\ \bibnamefont
  {Verbridge}}, \bibinfo {author} {\bibfnamefont {D.~R.}\ \bibnamefont
  {Southworth}}, \bibinfo {author} {\bibfnamefont {B.}~\bibnamefont {Ilic}},
  \bibinfo {author} {\bibfnamefont {H.~G.}\ \bibnamefont {Craighead}}, \ and\
  \bibinfo {author} {\bibfnamefont {J.~M.}\ \bibnamefont {Parpia}},\ }\bibfield
   {title} {\enquote {\bibinfo {title} {High-$q$ nanomechanics via destructive
  interference of elastic waves},}\ }\href {\doibase
  10.1103/PhysRevLett.106.047205} {\bibfield  {journal} {\bibinfo  {journal}
  {Phys. Rev. Lett.}\ }\textbf {\bibinfo {volume} {106}},\ \bibinfo {pages}
  {047205} (\bibinfo {year} {2011})}\BibitemShut {NoStop}%
\bibitem [{\citenamefont {Roukes}(2001)}]{Roukes2001}%
  \BibitemOpen
  \bibfield  {author} {\bibinfo {author} {\bibfnamefont {M.~L.}\ \bibnamefont
  {Roukes}},\ }\bibfield  {title} {\enquote {\bibinfo {title}
  {Nanoelectromechanical systems},}\ }in\ \href@noop {} {\emph {\bibinfo
  {booktitle} {Transducers '01 Eurosensors XV}}},\ \bibinfo {editor} {edited
  by\ \bibinfo {editor} {\bibfnamefont {E.}~\bibnamefont {Obermeier}}}\
  (\bibinfo  {publisher} {Springer Berlin Heidelberg},\ \bibinfo {address}
  {Berlin, Heidelberg},\ \bibinfo {year} {2001})\ pp.\ \bibinfo {pages}
  {658--661}\BibitemShut {NoStop}%
\bibitem [{\citenamefont {Schmid}\ \emph {et~al.}(2013)\citenamefont {Schmid},
  \citenamefont {Kurek}, \citenamefont {Adolphsen},\ and\ \citenamefont
  {Boisen}}]{Schmid2013}%
  \BibitemOpen
  \bibfield  {author} {\bibinfo {author} {\bibfnamefont {S.}~\bibnamefont
  {Schmid}}, \bibinfo {author} {\bibfnamefont {M.}~\bibnamefont {Kurek}},
  \bibinfo {author} {\bibfnamefont {J.~Q.}\ \bibnamefont {Adolphsen}}, \ and\
  \bibinfo {author} {\bibfnamefont {A.}~\bibnamefont {Boisen}},\ }\bibfield
  {title} {\enquote {\bibinfo {title} {Real-time single airborne nanoparticle
  detection with nanomechanical resonant filter-fiber},}\ }\href {\doibase
  10.1038/srep01288} {\bibfield  {journal} {\bibinfo  {journal} {Scientific
  Reports}\ }\textbf {\bibinfo {volume} {3}},\ \bibinfo {pages} {1--5}
  (\bibinfo {year} {2013})},\ \Eprint
  {http://arxiv.org/abs/https://doi.org/10.1038/srep01288}
  {https://doi.org/10.1038/srep01288} \BibitemShut {NoStop}%
\bibitem [{\citenamefont {Hanay}\ \emph {et~al.}(2012)\citenamefont {Hanay},
  \citenamefont {Kelber}, \citenamefont {Naik}, \citenamefont {Chi},
  \citenamefont {Hentz}, \citenamefont {Bullard}, \citenamefont {Colinet},
  \citenamefont {Duraffourg},\ and\ \citenamefont {Roukes}}]{Hanay2012}%
  \BibitemOpen
  \bibfield  {author} {\bibinfo {author} {\bibfnamefont {M.~S.}\ \bibnamefont
  {Hanay}}, \bibinfo {author} {\bibfnamefont {S.}~\bibnamefont {Kelber}},
  \bibinfo {author} {\bibfnamefont {A.~K.}\ \bibnamefont {Naik}}, \bibinfo
  {author} {\bibfnamefont {D.}~\bibnamefont {Chi}}, \bibinfo {author}
  {\bibfnamefont {S.}~\bibnamefont {Hentz}}, \bibinfo {author} {\bibfnamefont
  {E.~C.}\ \bibnamefont {Bullard}}, \bibinfo {author} {\bibfnamefont
  {E.}~\bibnamefont {Colinet}}, \bibinfo {author} {\bibfnamefont
  {L.}~\bibnamefont {Duraffourg}}, \ and\ \bibinfo {author} {\bibfnamefont
  {M.~L.}\ \bibnamefont {Roukes}},\ }\bibfield  {title} {\enquote {\bibinfo
  {title} {Single-protein nanomechanical mass spectrometry in real time},}\
  }\href {\doibase 10.1038/nnano.2012.119} {\bibfield  {journal} {\bibinfo
  {journal} {Nature Nanotechnology}\ }\textbf {\bibinfo {volume} {7}},\
  \bibinfo {pages} {602--608} (\bibinfo {year} {2012})},\ \Eprint
  {http://arxiv.org/abs/https://doi.org/10.1038/nnano.2012.119}
  {https://doi.org/10.1038/nnano.2012.119} \BibitemShut {NoStop}%
\bibitem [{\citenamefont {Wilson-Rae}(2008)}]{Wilson-Rae2008}%
  \BibitemOpen
  \bibfield  {author} {\bibinfo {author} {\bibfnamefont {I.}~\bibnamefont
  {Wilson-Rae}},\ }\bibfield  {title} {\enquote {\bibinfo {title} {Intrinsic
  dissipation in nanomechanical resonators due to phonon tunneling},}\ }\href
  {\doibase 10.1103/PhysRevB.77.245418} {\bibfield  {journal} {\bibinfo
  {journal} {Phys. Rev. B}\ }\textbf {\bibinfo {volume} {77}},\ \bibinfo
  {pages} {245418} (\bibinfo {year} {2008})}\BibitemShut {NoStop}%
\bibitem [{\citenamefont {Moura}(2019)}]{Moura2019}%
  \BibitemOpen
  \bibfield  {author} {\bibinfo {author} {\bibfnamefont {J.~P.}\ \bibnamefont
  {Moura}},\ }\emph {\bibinfo {title} {Making light jump}},\ \href@noop {}
  {Ph.D. thesis},\ \bibinfo  {school} {Delft University of Technology}
  (\bibinfo {year} {2019})\BibitemShut {NoStop}%
\bibitem [{\citenamefont {Tao}\ \emph {et~al.}(2015)\citenamefont {Tao},
  \citenamefont {Navaretti}, \citenamefont {Hauert}, \citenamefont {Grob},
  \citenamefont {Poggio},\ and\ \citenamefont {Degen}}]{Tao2015}%
  \BibitemOpen
  \bibfield  {author} {\bibinfo {author} {\bibfnamefont {Y.}~\bibnamefont
  {Tao}}, \bibinfo {author} {\bibfnamefont {P.}~\bibnamefont {Navaretti}},
  \bibinfo {author} {\bibfnamefont {R.}~\bibnamefont {Hauert}}, \bibinfo
  {author} {\bibfnamefont {U.}~\bibnamefont {Grob}}, \bibinfo {author}
  {\bibfnamefont {M.}~\bibnamefont {Poggio}}, \ and\ \bibinfo {author}
  {\bibfnamefont {C.~L.}\ \bibnamefont {Degen}},\ }\bibfield  {title} {\enquote
  {\bibinfo {title} {Permanent reduction of dissipation in nanomechanical si
  resonators by chemical surface protection},}\ }\href {\doibase
  10.1088/0957-4484/26/46/465501} {\bibfield  {journal} {\bibinfo  {journal}
  {Nanotechnology}\ }\textbf {\bibinfo {volume} {26}},\ \bibinfo {pages}
  {465501} (\bibinfo {year} {2015})}\BibitemShut {NoStop}%
\bibitem [{\citenamefont {Brunet}\ \emph {et~al.}(2017)\citenamefont {Brunet},
  \citenamefont {Aureau}, \citenamefont {Chantraine}, \citenamefont
  {Guillemot}, \citenamefont {Etcheberry}, \citenamefont {Gouget-Laemmel},\
  and\ \citenamefont {Ozanam}}]{Brunet2017}%
  \BibitemOpen
  \bibfield  {author} {\bibinfo {author} {\bibfnamefont {M.}~\bibnamefont
  {Brunet}}, \bibinfo {author} {\bibfnamefont {D.}~\bibnamefont {Aureau}},
  \bibinfo {author} {\bibfnamefont {P.}~\bibnamefont {Chantraine}}, \bibinfo
  {author} {\bibfnamefont {F.}~\bibnamefont {Guillemot}}, \bibinfo {author}
  {\bibfnamefont {A.}~\bibnamefont {Etcheberry}}, \bibinfo {author}
  {\bibfnamefont {A.~C.}\ \bibnamefont {Gouget-Laemmel}}, \ and\ \bibinfo
  {author} {\bibfnamefont {F.}~\bibnamefont {Ozanam}},\ }\bibfield  {title}
  {\enquote {\bibinfo {title} {Etching and chemical control of the silicon
  nitride surface},}\ }\href {\doibase 10.1021/acsami.6b12880} {\bibfield
  {journal} {\bibinfo  {journal} {ACS Applied Materials \& Interfaces}\
  }\textbf {\bibinfo {volume} {9}},\ \bibinfo {pages} {3075--3084} (\bibinfo
  {year} {2017})},\ \bibinfo {note} {pMID: 27977928},\ \Eprint
  {http://arxiv.org/abs/https://doi.org/10.1021/acsami.6b12880}
  {https://doi.org/10.1021/acsami.6b12880} \BibitemShut {NoStop}%
\end{thebibliography}%
	
\end{document}